# Effectiveness of Ninth-Grade Physics in Maine: Conceptual Understanding

Michael J. O'Brien, Kennebunk High School, Kennebunk, ME
John R. Thompson, The University of Maine, Orono, ME

The *Physics First* movement—teaching a true physics course to $9^{th}$ grade students—is gaining popularity in high schools. There are several different rhetorical arguments for and against this movement, and it is quite controversial in physics education. However, there is no actual evidence to assess the success, or failure, of this substantial shift in the science teaching sequence. We have undertaken a comparison study of physics classes taught in $9^{th}$- and $12^{th}$ grade classes in Maine. Comparisons of student understanding and gains with respect to mechanics concepts were made with excerpts from well-known multiple-choice surveys and individual student interviews. Results indicate that both populations begin physics courses with similar content knowledge and specific difficulties, but that in the learning of the concepts $9^{th}$ graders are more sensitive to the instructional method used.

**Background**

Advocates of the *Physics First* movement[1-3] argue that the current sequence of high school science courses (biology-chemistry-physics) should be flipped in response to the dramatic changes in science curricula during the $20^{th}$ century.[2] Because of the important discoveries in biology that have been made since Watson and Crick's discovery of the structure of DNA in 1953, modern biology courses emphasize molecular methods, genetics and biochemistry; this is very different from the general biology classes of the early 20$^{th}$ century that were a composite of botany, physiology, and zoology.  Chemistry courses have undergone a similar evolution:  modern chemistry emphasizes atomic structure, atomic energy levels, and even some quantum theory. *Physics First* advocates argue that a good understanding of modern biology requires a chemistry background, and a good understanding of modern chemistry requires a physics background.

Beyond these changes that have occurred in the content of high school science courses, advocates for *Physics First* cite other potential advantages to teaching physics to 9$^{th}$ graders rather than 12$^{th}$ graders:[3]

1. Improved learning of algebra due to concurrent applications of the algebra to physics.

2. Increased enrollment in physics courses.  Currently only about 35% of high school students take physics, and approximately 25% of high school students take both chemistry and physics.

3. The ability for students to take advanced or elective science courses during 12$^{th}$ grade without having to take two science courses simultaneously.

4. A better foundation of science content and skills (forces, motion, energy, experimental design, and data analysis) than earth science.

5. Improved integration of topics between biology, chemistry and physics, due to the increasing conceptual grain size of the topics.

**Previous Research into the Effectiveness of *Physics First***

While these arguments presented by *Physics First* advocates may seem logical, there is a scarcity of empirical data that could help determine the extent to which a *Physics First* program actually benefits students. Many educators have reported success in teaching physics first, but these reports have been mostly anecdotal and lacking quantitative data. There have been small-scale studies of the effectiveness of teaching physics to underclassmen (9th and 10th graders) that have been published since the *Physics First* movement's infancy in the late 1960's.[4-8] However, the teaching situations in most of these studies do not reflect the classroom situation for the vision of *Physics First,* namely classes for all 9th grade students. Thus we feel that these studies lack the necessary generalizability to be valid evidence of effectiveness.

Dreon[9] described the state of *Physics First* curricula in 13 public and private Pennsylvania high schools. Dreon's study contains no discussion of student learning in these courses. However, he states that an overview of content and context for these courses allows us to "move one step closer to answering what [he] believe[s] to be one of the most important questions facing the Physics First movement: 'Can ninth-grade students successfully learn physics?'"

Korsunsky and Agar[10] recently reported results of a survey on student attitudes and expectations among 8th graders. They document widely varying results, "distorted" expectations about the 9th grade physics course they would be taking the following year, and gender differences in student perceptions of physics. They note that there are "virtually no reports of actual research studies relevant to Physics First."

Most recently, Goodman and Etkina[11] reported the results of teaching a "mathematically rigorous" 9th-grade physics course using algebra but not trigonometry, whose content is derived from the AP Physics B curriculum. The course was implemented with great success, as measured by an increase in the number of students taking and passing the AP exams and a comparison of AP B performance with the 1998 TIMMS scores. The extent to which the AP exam serves as a valid measure of *conceptual* understanding is debatable; nevertheless, this result serves as a pragmatic indicator of student success with a widely recognized assessment instrument (the AP exam). The success of this mathematically rigorous method of instruction in the Rutgers study is consistent with work in introductory college courses integrating calculus and physics.[12]

On a much larger scale, Sadler and Tai[13] conducted a study of the effect of high school science and mathematics on grades in college introductory science courses, with ~8500 students at 63 colleges and universities. Their results indicate that high-school science courses are not associated with better performance in introductory college science courses "out-of-discipline"; for example, taking high school physics is associated with higher introductory college physics grades, but not with grades in introductory college chemistry or biology. They also find that the number of years of high-school math taken correlates with higher college science grades across the board. They suggest that these results can be extrapolated to the argument that taking physics in 9th grade will improve chemistry learning in 10th grade, and similarly for 11th grade biology. They state that their result "casts doubt on the impact of changing the traditional high-school science sequence" to physics first. However, the transition from high school to college courses is very different from the ninth- to 10th-grade transition, so it's not clear how directly these results can be applied to the Physics First situation.

Our study directly investigates $9^{th}$ grade physics students and documents the content understanding and gains in $9^{th}$-grade physics classes and the attitudes and expectations of students in these classes, using well researched survey questions with known outcomes for different instructional methods.

**Our Study**

We have chosen to compare the experiences of $9^{th}$ graders taking physics and $12^{th}$ graders taking physics for the first time. The research question we address in this paper is whether there is a difference in the performance of $9^{th}$ graders and $12^{th}$ graders on a survey of kinematics and mechanics concepts.

The intended study population is typical high school students in the state of Maine. Seven high schools in Maine participated in this study, providing a total of 321 students. Three of the schools teach physics to ninth graders, and three teach physics to twelfth graders. One of the participating schools teaches physics to ninth graders and also has a course for twelfth graders who did not take physics in an earlier grade. The participating schools are schools that responded to a request sent out on the Maine Science Listserv, and volunteered to participate.

Because of the different levels of students within the participating schools, and the different teaching methods employed by the different schools, five distinct groups of students exist in this study:

1. Ninth-grade students who experienced traditional instruction. (n=80)

2. Ninth-grade students who experienced Modeling-based instruction (as described by Wells, Hestenes, and Swackhamer[14]). (n=32)

3. Ninth-grade honors-level students who experienced traditional instruction. (n=28)

4. Ninth-grade honors-level students who experienced Modeling-based instruction. (n=76)

5. Twelfth-grade students who experienced traditional instruction. (n=105)

All of the 9th grade courses studied here used Hewitt's textbook[15] as a basis for the course, allowing for some measure of control of that variable.  This also means that we can expect our data to be representative of the majority of physics first courses:  eight of the 13 Pennsylvania courses surveyed by Dreon[9] used Hewitt's book.  (Data from the American Institute of Physics indicate that Hewitt is used in 83% of introductory college-level physics courses for non-science majors.[16])  Of course the extent to which the Modeling courses utilize any text is open to debate, but nevertheless, this is what the teachers reported.

Note that all of the twelfth-grade students were non-honors level whose teachers employed a traditional method of instruction.  The inclusion of sub-groups of twelfth grade students at the honors level and sub-groups who experienced Modeling-based instruction would offer a more complete comparison of the 9th- and 12th grade populations in this study.

*Physics First* is a relatively new paradigm, and one that has not been tested empirically.  When we conducted this study, there were no instruments designed specifically to assess 9th grade physics students.  (Researchers from Arizona State

University have recently developed a linguistically simplified version of the *Force Concept Inventory* for 9th graders that is easier to read.[17]) We wished to create an instrument that would be appropriate for 9th graders, and whose results could be compared to previous studies of student learning of mechanics concepts.  For this reason, excerpts from established instruments were used to create an instrument judged (by the authors) to be appropriate for assessing 9th graders.  We used excerpts from three different and well established instruments—*The Force Concept Inventory* (FCI),[18] *The Force and Motion Conceptual Evaluation* (FMCE),[19] and *The Test for Understanding Graphs in Kinematics* (TUG-K)[20]—to assemble a 27-question multiple-choice survey that was used to evaluate students' understanding of mechanics concepts, particularly kinematics graphs and Newton's Laws.

This survey was given to students as a pretest in September 2005, and as a post-test sometime after instruction in mechanics was completed.  Depending on the duration of the mechanics instruction of the different classes, the post-test was administered sometime between December 2005 and March 2006.

This report presents data from part of a larger study that included surveys of attitudes and expectations as well.  We will include that data and its implications in a future publication.

**Results and Discussion**

On the pretest, both the 9th and 12th graders did only slightly better than random guessing:  while random guessing would result in answering 4.4 questions out of 27 correctly (16%), the ninth graders answered 5.3 questions correctly (20%), and the twelfth graders answered 6.0 questions correctly (22%) (Table I).  (We note that the pretest results for the different subpopulations in 9$^{th}$ grade were sufficiently similar to warrant a single value for all 9$^{th}$ grade students.)  This indicates that students from both grade levels had very little conceptual understanding of mechanics at the beginning of their physics courses.  These pretest scores are similar to those of 12$^{th}$ graders on the complete versions of the FCI and FMCE.[18,19]

|  | 9$^{th}$ Grade | 12$^{th}$ Grade |
|---|---|---|
| N | 216 | 105 |
| Mean Score (out of 27) | 5.3 | 6.0 |
| % Correct | 20 | 22 |

**Table I:  Overall Mechanics Concept Survey Pretest Scores**

We used normalized gain *(<g> = (post-test score – pretest score) / (perfect score – pretest score))* as a means of measuring conceptual learning, since it is commonly used as a figure of merit for instruction measured by the FCI and the FMCE.  On the post-test, results between sub-groups differed, often substantially.  The results for each sub-group are displayed in Table II.  The honors-level 9$^{th}$ graders had the highest post-test scores and normalized gains of any of the sub-groups, even above that of the 12$^{th}$ grade classes.  The non-honors-level 9$^{th}$ graders had the lowest post-test scores and normalized gains of any of the sub-groups.  All of the normalized gains are significant except for those of the non-honors-level 9$^{th}$ graders that did not receive Modeling-based instruction.  For this sub-group only, the post-test scores are not significantly different from the pretest scores.

| Sub-group | | N | Pretest Score (out of 27) | Post-test Score (out of 27) | <g> | p-value (post-test vs. pretest) |
|---|---|---|---|---|---|---|
| Grade | Honors (H/N) | Modeling (M/N) | | | | |
| 9 | N | N | 80 | 5.6 | 6.3 | 3% | 0.072 |
| 9 | N | M | 32 | 5.0 | 8.9 | 18% | 0.000 |
| 9 | H | N | 28 | 5.5 | 13.0 | 35% | 0.000 |
| 9 | H | M | 76 | 4.9 | 12.5 | 35% | 0.000 |
| 12 | N | N | 105 | 6.0 | 10.9 | 23% | 0.000 |

**Table II:  Overall post-test scores and normalized gains (<g>) broken down by grade, course, and instructional method**

The use of Modeling Instruction[14] appeared to have a large effect on the non-honors-level, ninth-grade students' performance on the post-test and their normalized gains.  These results are not inconsistent with data that have been collected on the efficacy of Modeling in previous studies.  Hake,[21] Wells et al.,[14] and Hestenes[22] found that the normalized gains of students—on the full FCI—in courses that employ Modeling are approximately twice as large (40-60%) as those of students in traditionally taught courses (20-35%).  This was generally true for both honors and non-honors level students.  However, in our study, there was not a significant difference in the normalized gains of the two honors groups (Modeling vs. traditional.)  The honors groups outperformed the non-honors groups, regardless of the type of instruction.

An interesting side note about the use of Modeling-based instruction is that the use of this method demands more instructional time.  The effect on the normalized gains of the number of weeks spent on mechanics instruction was found to be not significant using ANOVA.  This indicates that the interactive method of instruction, and not the amount of traditional instruction, is the more important variable in student learning.  This is consistent with previous research at many levels.[21]

**Implications and Limitations**

The similarity of the pretest results between the ninth- and 12th-graders with respect to their conceptual understanding of physics does not rule out the feasibility of teaching ninth-grade physics.

It is important to keep in mind that ninth-grade *Physics First* classes are taught to *all* ninth-graders, while a 12th-grade course is taken as an elective by a subset of the students in 12th grade.  Since the majority of ninth-graders taking physics are regular students, the results for these students are the most relevant.

The relative success of the non-honors-level ninth graders whose teachers employed a modeling-based method of instruction compared to their counterparts whose teachers did not use modeling indicates that schools and teachers considering teaching physics to these students need to carefully consider how the course will be taught.  These results suggest that in order for these students to be able to understand the basic kinematics and mechanics concepts that are typically taught in an introductory high school physics course, teachers need to employ a more student-centered approach rather than the traditional lecture-based approach that is employed in most 12th-grade courses.

As Art Hobson says in the November 2005 edition of *The Physics Teacher,* "*Physics First* will succeed or fail depending on the way it is implemented.  If all it does is offer a math-based first course focusing on classical physics, similar to many first physics courses now offered in the 11th or 12th grade, it will fail for the same reason that those courses fail."[24]

An alternate interpretation of the data stems from the observation that the nonmodeling, regular 12th-grade students outperformed the regular ninth-grade students regardless of instruction type, implying that a conceptual understanding would be better achieved by a 12th-grade physics course than one in ninth grade. However, this interpretation neglects the elective nature of the 12th-grade course, suggesting a self-selection mechanism.

We recognize that there are limits to the extent to which our own results can be generalized. First, our sample size is somewhat small. Data from more students at more schools would help limit the influence of variables such as differences in curricula, teaching styles, and student populations that may exist from school to school. A broader selection of schools may also address an issue of self-selection by the teachers who volunteered for the study. Second, all of the twelfth-grade students in this study were at the non-honors level taught with a traditional method of instruction. It would be helpful to include sub-groups of $12^{th}$-graders at the honors level and sub-groups who received modeling-based instruction to have better control groups between the two grades. Obtaining data from schools that require physics of all $12^{th}$-graders would provide an excellent control for the selectivity factor.

For the *Physics First* movement to be an evidence-based educational movement with data to support its justifications, a data corpus must be built that includes larger scale studies on student learning in ninth-grade physics courses, in addition to studies that follow students into subsequent chemistry and biology courses and perform analogous studies in those courses as well as longitudinal studies.

Nevertheless, our study is among the first steps toward a scholarly analysis of this movement. Our results yield insight into the effectiveness of *Physics First* at improving conceptual understanding of mechanics. However, until more empirical data are available, the academic value of a *Physics First* curriculum will be merely a matter of opinion.

## Acknowledgments

We are grateful for helpful comments from an anonymous reviewer.  We acknowledge partial support for this work from the Maine Economic Improvement Fund and for the preparation of this manuscript from the Maine Academic Prominence Initiative.

O'Brien and Thompson	Ninth Grade Physics in Maine:  Conceptual Understanding12. K. Marrongelle, K. Black, and D. Meredith, "Studio Calculus Physics: Interdisciplinary Mathematics with Active Learning," from *Integrated Mathematics: Choices and Challenges* (Sue Ann McGraw, National Council of Teachers of Mathematics, 2003).

13. Philip M. Sadler and Robert H. Tai, "The Two High-School Pillars, Supporting College Science," *Science* 317, 457-458 (2007).

14. M. Wells, D. Hestenes, and G. Swackhamer, "A Modeling Method for High School Physics Instruction," *Am. J. Phys.* 63 (7), 606-619 (1995).

15. Hewitt, P.G., *Conceptual Physics,* 10th ed. (Pearson Prentice Hall, Upper Saddle River, NJ, 2005).

16. M. Neuschatz, and M. McFarling, "Broadening the Base:  High School Physics Education at the turn of a New Century" (2003).  Downloaded from http://www.aip.org/statistics/trends/highlite/hs2001/hshigh.htm.  Last accessed November 11, 2007.

17. J. Jackson, personal communication.

18. D. Hestenes, M. Wells, and G. Swackhamer, "Force Concept Inventory," *Phys. Teach.* 30, 141-158 (March 1992).

19. R.K. Thornton and D.R. Sokoloff, "Assessing Student Learning of Newton's Laws: The Force and Motion Conceptual Evaluation and Evaluation of Active Learning Laboratory and Lecture Curricula," *American Journal of Physics* 66, 338-352 (1998).

20. R.J. Beichner, "Testing Student Interpretation of Kinematics Graphs," *Am. J. Phys.* 62 (8), 750-762 (1994).
14